# On the Performance of DCSK MIMO Relay Cooperative Diversity in Nakagami-*m* and Generalized Gaussian Noise Scenarios

Ehab Salahat, *Student Member*, IEEE
`ehab.salahat(at)ieee(dot)org`

*Abstract*—Chaotic Communications have drawn a great deal of attention to the wireless communication industry and research due to its limitless meritorious features, including excellent anti-fading and anti-intercept capabilities and jamming resistance *exempli gratia*. Differential Chaos Shift Keying (DCSK) is of particular interest due to its low-complexity and low-power and many attractive properties. However, most of the DCSK studies reported in the literature considered the additive white Gaussian noise environment in non-cooperative scenarios. Moreover, the analytical derivations and evaluation of the error rates and other performance metrics are generally left in an integral form and evaluated using numerical techniques. To circumvent on these issues, this work is dedicated to present a new approximate error rates analysis of multi-access multiple-input multiple-output dual-hop relaying DCSK cooperative diversity (DCSK-CD) in Nakagami-*m* fading channels (enclosing the Rayleigh fading as a particular case). Based on this approximation, closed-form expressions for the average error rates are derived for multiple relaying protocols, namely the error-free and the decode-and-forward relaying. Testing results validate the accuracy of the derived analytical expressions.

*Keywords*—DCSK, Multi-Access, Relay, Cooperative Diversity, Chaos Communication, Bit Error Rates, Nakagami-m, Rayleigh.

## I. INTRODUCTION

CHAOTIC MODULATIONS have drawn a great deal of attention in the wireless communications industry and research community [1]. Chaotic signals are non-periodic random-like signals that are generated from nonlinear dynamic systems. They are naturally well-suited for spread spectrum wireless communications due to their inherently wideband characteristics [2]. In contrast to the classical direct sequence spread spectrum (DSSS) communications that use pseudo-random noise (PN) sequences, chaos communications utilize chaotic sequences which are directly generated by a discrete-time nonlinear map (e.g., logistic map) [1]. They possess a myriad of merits, in addition to those accomplished by classical DSSS, including mitigating frequency selectivity, jamming-resistance, low probability of interception, easiness of generation, simple transceiver circuits, strong immunity to self-interference, anti-multipath, transmission security, and excellent cross-correlation properties [3]. These features make it an excellent candidate for military scenarios, ultra-dense populated environments, and many other applications [4] [5]. Chaos-based sequences were also proven to reduce the peak-to-average power ratio (PAPR) [5].

Since its introduction in [6] by Parlitz *et al.*, chaotic digital modulation, a great deal of research effort has been devoted to developing new chaotic modulation schemes. Depending on the need for chaotic synchronization, chaotic modulation schemes can be categorized into coherent and non-coherent schemes [4]. Amongst them, the coherent chaos-shift-keying (CSK) and non-coherent differential CSK (DCSK) are the prominent ones. Coherent CSK, akin to coherent modulations, requires perfect knowledge of channel state information and the chaotic synchronization are required to generate a chaotic replica at the receiver side to perform the demodulation [7], and hence are not really easily implementable in fast-fading channels with short coherence time [3]. For example, the proposed chaotic synchronization in [8] is still practically impossible to achieve in noisy environments. Hence, and due to these limitations and challenges, non-coherent chaos-based communications are more promising, and amid them, DCSK is the most attractive one [9].

DCSK, similar to differential phase shift keying (DPSK), is a non-coherent scheme, and so requires no synchronization or channel state information to recover the transmitted messages. However, DCSK is more robust to multipath fading compared than DPSK schemes [7] [10]. DCSK was also extended to the non-binary domain, leading to *M*-ary DCSK [11]. Currently, DCSK was considered for low-power and low-complexity wireless applications such as wireless personal area networks (WPANs) and wireless sensor networks (WSNs). It was also considered recently for short-range ultra-wideband (UWB) communications [1]. An elegant survey that studies recent advances in DCSK and its derivatives (e.g. RM-DCSK [2] a d HE-DCSK [12]) is given in [1] for the reader's reference.

Motivated by the advantages of DCSK, significant amount of research work was carried to study and analyze the theoretical performance and fundamental limits of the DCSK systems under different channel conditions such as the additive white Gaussian noise (AWGN) and Rayleigh fading channels with single-input single-output (SISO) links. However, in realistic and practical communication scenarios, the transmitters may not be able to use multiple antennas due to size, complexity, power, cost and other constrains. To tackle this design issue, a novel MIMO relay DCSK cooperative-diversity system that used a single-transmit antenna, multi-antenna relay and multi-antenna receiver, was analyzed in [13]. The different links in the design were assumed to be subjected to Nakagami-*m* fading and AWGN channels. The authors derived a closed-form analytical expression that was given in an integral form, which required numerical evaluation due to its complexity. Moreover, the AWGN noise assumption may sometimes be inaccurate and unsuitable to model the noise conditions [2].

To circumvent on these issues, and inspired by [13], this paper is dedicated to present a novel performance analysis of the same MIMO relay DCSK cooperative-diversity proposed by the authors of [13], with the assumption of an additive white generalized Gaussian noise (AWGGN) instead. Moreover, to simplify the analytical derivations in the error rate analysis, we utilize an approximated analysis approach in our modeling and mathematical derivations. Numerical evaluations show an agreement with the approximated expressions.

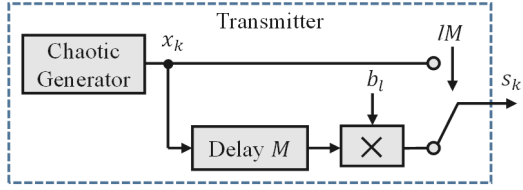

Fig. 1: Block diagram of DCSK transmitter [14].

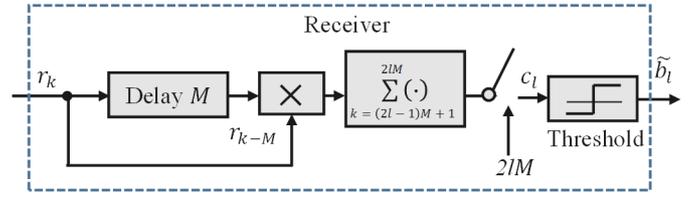

Fig. 2: Block diagram of DCSK receiver [14].

The rest of this paper is given as follows. In section II, a brief overview of the DCSK and the deployed cooperative scheme is presented. The analytical derivation of the new DCSK average error rates approximation is developed in section III. Analytical results in different fading and noise scenarios and different configurations of the deployed cooperative scheme are presented and compared to numerical evaluations, which show an excellent agreement. Finally, the paper findings are summarized in section IV.

## II. SYSTEM MODEL

### A. Principle of DCSK Modulation

The designs of the transmitter and the receiver of the DCSK system are shown in Fig. 1 and Fig. 2, respectively. In the DCSK, the $l^{th}$ transmitted symbol is assumed to be either $+1$ or $-1$, with equal probabilities. During the $l^{th}$ transmission, the transmitted signal, $s_k$, is given by:

$$s_k = \begin{cases} x_k, & k = 2(l-1)M+1, \dots, (2l-1)M \\ b_l x_{k-M}, & k = (2l-1)M+1, \dots, 2lM \end{cases}, \quad (1)$$

where $2M$ is the spreading factor. At the receiver, the received signal $r_k$, that is the input signal to the correlator, is given by:

$$r_k = \varphi s_k + N. \quad (2)$$

where $\varphi$ is the fading parameter and $N$ is the additive noise to the signal. The $l^{th}$ decoded symbol, $c_l$, is compared to threshold 0 to decide $\tilde{b}_l$. The decision and hence the recovered symbol $\tilde{b}_l$ is given by:

$$\tilde{b}_l = \begin{cases} +1, & c_l \geq 0 \\ -1, & c_l < 0 \end{cases}. \quad (3)$$

Referring to [14], and making use of the relation between the error function the $Q$-function [15], the BER of the decoding data bits is given by

$$\text{BER}_{\text{DCSK}} = \tfrac{1}{2}\text{erfc}\left(\tfrac{4}{\gamma}\left[1+\tfrac{M}{2\gamma}\right]^{-\tfrac{1}{2}}\right) = Q\left(\sqrt{\tfrac{\gamma^2}{2\gamma+M}}\right), \quad (4)$$

where $\gamma$ is the signal-to-noise ratio, which is exactly the same BER performance expression of the DDCSK-Walsh Coding reported in [16], and hence, the results derived in the later sections apply for both DCSK and DDCSK-Walsh Coding.

### B. Multi-Access MIMO Relay DCSK-CD

Consider a two-hop network system model with $n$ users (sources), one multi-antenna destination and a multi-antenna relay. We assume that each transmitter has a single antenna, and that there is no cooperation between users to transmit messages. As such, only the multi-antenna relay helps in transmitting messages to the final receiver. Furthermore, assume that a transmission time is divided into two phases (time-slots), the broadcast and cooperate, respectively. In the broadcast phase, users broadcast their messages to other terminals (the relay and the receiver). In the cooperate phase, the relay cooperates with the users to send their messages to the destination [2] [13]. The system model, assuming 2 users for demonstration only, is shown in Fig. 3, where each user can support a single transmit antenna, and the relay has $M_R$ antennas used under ideal condition (error free (EF) at relay antennas) or decode-and-forward (DF) protocol, and the destination has $M_D$ receive antennas [2] [13].

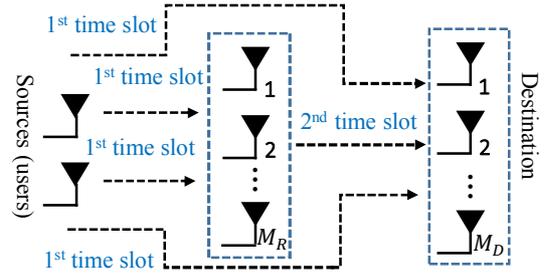

Fig. 3: System model for 2-user MIMO relay DCSK-CD [2] [13].

As mentioned earlier, we will assume that the fading model of the individual links to be i.i.d. (independent and identically distributed) Nakagami-$m$ channels which are subjected to additive generalized Gaussian noise [17] [18], and the channel state remains constant during each transmission period. The probability density function for a gamma random variable (RV) is given by [19] [20]:

$$f(\gamma) = \frac{\gamma^{a-1}e^{-\gamma/b}}{b^a \Gamma(a)}, \quad (5)$$

where $a$ and $b$ are positive real numbers. We will adopt the notation $\mathcal{G}(a,b)$ for gamma RV henceforth. To carry the rest of the analysis, we shall recall the following two theorems that are related to gamma RVs [2] [13] [21].

***Theorem 1***: Given $N$ independent gamma RVs $X_1, X_2, \dots, X_N$, where $X_k \sim \mathcal{G}(a_k, b)$ $(k=1,\dots,N)$, the sum of these variables $X = \sum_{k=1}^{N} X_K$ is also a gamma RV, given as $X \sim \mathcal{G}(\sum_{k=1}^{N} a_k, b)$.

**Theorem 2**: Given $N$ independent gamma RVs $X_1, X_2, ..., X_N$, where $X_k \sim \mathcal{G}(a_k, b_k)$ $(k = 1, ..., N)$ where $b_1 \neq b_2 \neq \cdots, \neq b_N$, then the sum of these variables $X = \sum_{k=1}^{N} X_K$ has a PDF that is expressed as:

$$f(x) = \sum_{i=0}^{\infty} \left[\frac{\eta_i C}{\Gamma(\rho+1)b_0^{\rho+1}}\right] x^{\rho+i-1} e^{-\frac{x}{b_0}}, \quad (6)$$

with $\eta_0 = 1$, $C = \prod_{k=1}^{N}\left(\frac{b_0}{b_k}\right)^{a_k}$, $\eta_{i+1} = \frac{1}{i+1}\sum_{t=1}^{i+1} t z_t \eta_{i+1-t}$, $i = 0,1,...$, $z_j = \sum_{k=1}^{N} a_k j^{-1}\left(1 - \frac{b_0}{b_k}\right)^j$, $j = 1,2,...$, $\rho = \sum_{k=1}^{N} a_k > 0$, and $b_0 = \min(b_k)$.

It can be consequently shown that, following the previous theorems and using the same approach that was used in [13] [19] [22], with an i.i.d fading channels assumption between any two terminals (source-relay (S-R), source-destination (S-D) and relay-destination (R-D)), the links are also given as gamma RVs in the forms:

$$\gamma_{SR} \sim \mathcal{G}\left(M_R mL, \frac{\tilde{\gamma}}{2M_R d_{SR}^2 mnL}\right) = \mathcal{G}(\alpha_{SR}, \beta_{SR}), \quad (7)$$

$$\gamma_{SD} \sim \mathcal{G}\left(M_D mL, \frac{\tilde{\gamma}}{2M_D d_{SD}^2 mnL}\right) = \mathcal{G}(\alpha_{SD}, \beta_{SD}), \quad (8)$$

$$\gamma_{RD} \sim \mathcal{G}\left(M_R M_D mL, \frac{\tilde{\gamma}}{2M_R M_D d_{RD}^2 mnL}\right) = \mathcal{G}(\alpha_{RD}, \beta_{RD}), \quad (9)$$

where $M_R$, $M_D$, $n$, $L$, $d_{SR}$, $d_{SD}$, $d_{RD}$, $\tilde{\gamma}$ and $m$ represent the number of deployed relay antenna, destination antenna, users served, and fading paths, and the distance between the S-R, S-D and R-D, the average signal-to-noise ratio, and the fading severity, respectively.

### III. AVERAGE ERROR RATES ANALYSIS

In this section, the analytical derivation of the average error rates of the proposed system model will be presented.

#### A. Novel Approximation of the Generalized Q-Function

It was shown in [14] that the BER of the DCSK is given as in (4). This expression is based on the Gaussian noise assumption and is very difficult to manipulate analytically. This Gaussian noise (additive white Gaussian noise) assumption therein might be valid in some realistic and practical communications scenarios. Hence, in this work and as was done in [2], we extend further the Gaussian assumption, given in terms of the Gaussian $Q$-function, to the generalized Gaussian noise model, given in terms of the generalized $Q$-function, which is written as [17]:

$$Q_a(x) = \frac{a\Lambda_0^{2/a}}{2\Gamma(1/a)}\int_x^{\infty} e^{-\Lambda_0^a |u|^a} du = \frac{\Lambda_0^{2/a-1}}{2\Gamma(1/a)}\Gamma(1/a, \Lambda_0^a |x|^a). \quad (10)$$

where $\Lambda_0 = \sqrt{\Gamma(3/a)/\Gamma(1/a)}$ and $\Gamma(\cdot)$ and $\Gamma(\cdot,\cdot)$ are the gamma and the incomplete gamma functions [23]. Table I illustrates how the noise special models can be achieved from (8).

TABLE I: RELATION BETWEEN $Q_a(x)$ AND SPECIAL NOISE MODELS.

| Noise Dist. | Impulsive | Laplacian | Gaussian | Uniform |
|---|---|---|---|---|
| $a$ | 0.0 | 1.0 | 2.0 | $\infty$ |

Furthermore, to simplify the ABER analysis, we propose to approximate (4), for a fixed value of $M$, as a sum of scaled decaying exponential functions, given as:

$$Q_a\left(\sqrt{\frac{\gamma^2}{2\gamma+M}}\right) \approx \sum_{r=1}^{4} \delta_r e^{-\mu_r \gamma}, \quad (11)$$

where the fitting parameters, $\delta_r$ and $\mu_r$ are obtained using nonlinear curve fitting (using the MATLAB® curve-fitting Marquardt-Levenberg algorithm), with sample fitting values for different cases of $a$ being presented in table II, assuming $M = 32$, a value commonly used in the literature.

TABLE II: FITTING PARAMETERS OF $Q_a(\sqrt{\cdot})$ APPROXIMATION

| $a$ | $\delta_1$ | $\delta_2$ | $\delta_3$ | $\delta_4$ | $\mu_1$ | $\mu_2$ | $\mu_3$ | $\mu_4$ |
|---|---|---|---|---|---|---|---|---|
| 1 | 0.1078 | 0.4294 | -0.009 | 0.1788 | 0.5424 | 0.2477 | 0.7834 | 0.1044 |
| 1.5 | 0.4140 | 0.0955 | 0.0928 | -0.127 | 0.2113 | 0.6214 | 0.1321 | 0.6197 |
| 2 | 0.2520 | 0.3976 | -0.611 | 0.4621 | 0.5162 | 0.2243 | 0.4096 | 0.2243 |
| 2.5 | 0.6083 | -1.1060 | 0.2360 | 0.8107 | 0.2541 | 0.3982 | 0.6922 | 0.2534 |

The relative absolute error plots of this approximation are given in Fig. 4. With simple and direct variable transform to the approx. in (9), $Q_a(\cdot)$ approx. is straightforward.

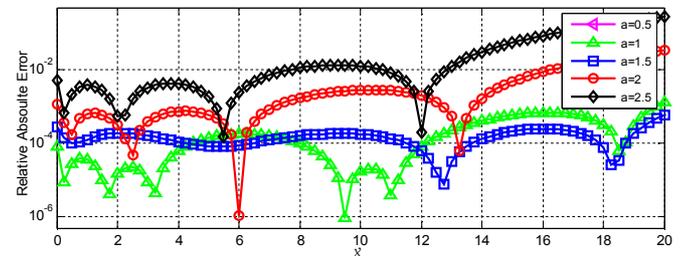

Fig. 4: Relative absolute Error for the approx. in (9).

#### B. ABER with Error Free Protocol for the Relay

The average bit error rates (ABER) due to a fading channel can be evaluated by averaging the conditional bit error rate probability of the noisy channel using the PDF of the fading envelope. With the aid of (4), the DCSK generalized error expression in AWGGN can be written using the averaging process expressed as [2]:

$$P_e = \int_0^{\infty} f_\gamma(\gamma) Q_a\left(\sqrt{\frac{\gamma^2}{2\gamma+M}}\right) d\gamma, \quad (12)$$

where $f(\gamma) = f(\gamma_D)$ for the error-free (EF) protocol is the fading PDF which, using theorems 1 and 2, is given at the destination as:

$$f(\gamma_D) = \psi \gamma_D^{\tilde{m}-1} e^{-\beta \gamma_D}, \quad (13.a)$$

with $\psi = \frac{\beta^{\tilde{m}}}{\Gamma(\tilde{m})}$, $\tilde{m} = \alpha_{SD} + \alpha_{RD}$, and $\beta = \sqrt{2}\beta_{SD}^{-1} = \sqrt{2}\beta_{RD}^{-1}$, or

$$f_\gamma(\gamma_D) = \sum_{i=0}^{\infty} \psi_i \gamma_D^{\tilde{m}-1} e^{-\beta \gamma_D}, \quad (13.b)$$

with $\psi_i = \left[\frac{\eta_i C}{\Gamma(\rho+1)b_0^{\rho+1}}\right]$, $\tilde{m} = \rho + i$, $\beta = \frac{1}{b_0}$, and $\beta_{SD} \neq \beta_{RD}$.

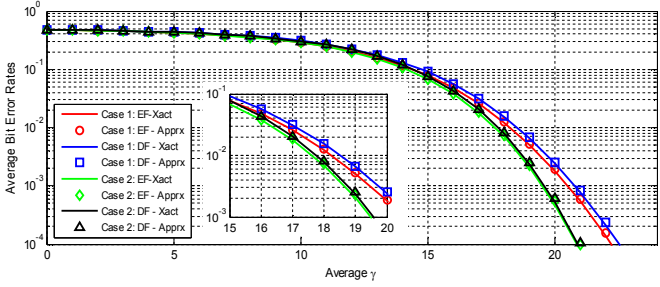

Fig. 5: test scenario 1 – ABER for EF and DF protocols in Rayleigh fading (case 1) and Nakagami-*m* fading (case 2) fading with *m*=4, and AWGN.

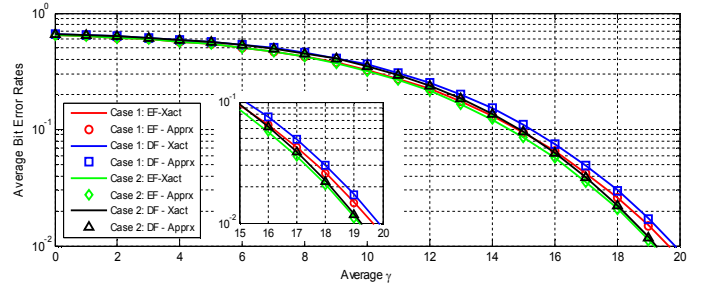

Fig. 6: test scenario 2 – ABER for EF and DF protocols in Rayleigh fading (case 1) and Nakagami-*m* fading (case 2) fading with *m*=4, and Laplacian noise.

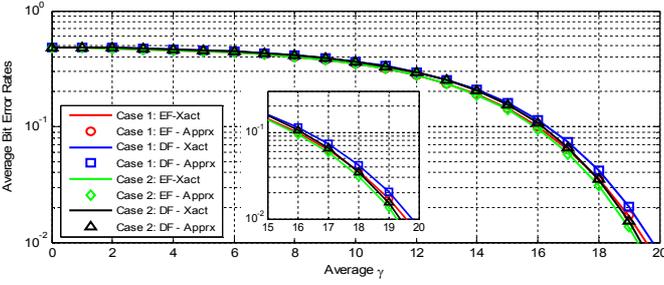

Fig. 7: test scenario 3 – ABER for EF and DF protocols in Rayleigh fading (case 1) and Nakagami-*m* fading (case 2) fading with *m*=4, and AWGN.

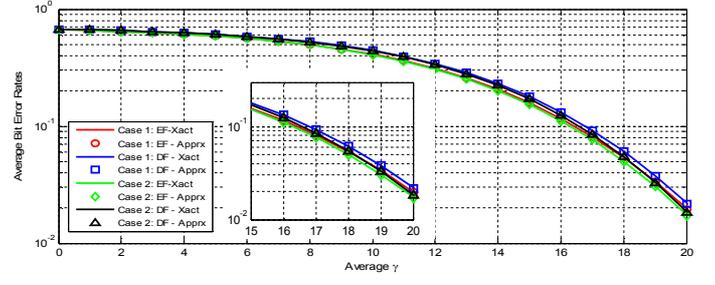

Fig. 8: test scenario 4 – ABER for EF and DF protocols in Rayleigh fading (case 1) and Nakagami-*m* fading (case 2) fading with *m*=4, and Laplacian noise.

Note that (12) is very difficult to be integrated and solved in a closed-form. Hence, the approx. in (11) is used to simplify the integrand. Using the approx. in (11) and the expression in (13.a), (12) can be evaluated as:

$$P_e \approx \sum_{r=1}^{4} \psi \delta_r \int_0^\infty \gamma^{\widetilde{m}-1} e^{-[\beta+\mu_r]\gamma} d\gamma = \sum_{r=1}^{4} \tilde{\psi}\Gamma(\widetilde{m}), \quad (14.a)$$

with $\tilde{\psi} = \frac{\psi \delta_r}{[\beta+\mu_r]^{\widetilde{m}}}$. Similarly, using (9) and (11.b), (10) can be solved as:

$$P_e = \sum_{i=0}^{\infty} \sum_{r=1}^{4} \psi_i \delta_r \int_0^\infty \gamma^{\widetilde{m}-1} e^{-[\beta+\mu_r]\gamma} d\gamma = \sum_{i=0}^{\infty} \sum_{r=1}^{4} \tilde{\psi}\Gamma(\widetilde{m}), \quad (14.b)$$

with $\tilde{\psi} = \frac{\psi_i \delta_r}{[\beta+\mu_r]^{\widetilde{m}}}$. These two expressions are new, analytically traceable and very easy to evaluate (take few seconds to be evaluated) using MATLAB®.

### C. ABER with Decode-and-Forward Protocol for the Relay

With the DF protocol, and following a similar approach to that of the EF protocol, the ABER can be obtained in a closed-form using the relation [2] [13] [24]:

$$\text{BER}_{\text{DF}} = \text{BER}_{\text{SR}} \cdot \text{BER}_{\text{SD}} + (1 - \text{BER}_{\text{SR}}) \cdot \text{BER}_{\text{D}}, \quad (15)$$

where each of the $\text{BER}_{\text{SR}}$ and $\text{BER}_{\text{SD}}$ are given exactly as in (14.a), while replacing the parameters $\{\widetilde{m}, \beta\}$ with $\{\alpha_{SR}, \beta_{SR}^{-1}\}$ and $\{\alpha_{SD}, \beta_{SD}^{-1}\}$, respectively, and $\text{BER}_{\text{D}}$ is exactly the same as the ABER of the EF protocol, of either form of (14.a) or (14.b) depending on (13.a) and (13.b), respectively. Hence (15) is finally given by:

$$\text{BER}_{\text{DF}} = \sum_{r_{SR}=1}^{4} \sum_{r_{SD}=1}^{4} \Psi_1 \Gamma(\widetilde{m}_{SR})\Gamma(\widetilde{m}_{SD}) + \sum_{r_D=1}^{4} \tilde{\psi}_D \Gamma(\widetilde{m}_D) - \sum_{r_D=1}^{4} \sum_{r_{SR}=1}^{4} \Psi_2 \Gamma(\widetilde{m}_D)\Gamma(\widetilde{m}_{SR}), \quad (16)$$

with $\Psi_1 = \tilde{\psi}_{SR}\tilde{\psi}_{SD}$ and $\Psi_2 = \tilde{\psi}_D \tilde{\psi}_{SR}$. This expression is novel and is the first to address multi-access MIMO DCSK-CD, which is also applicable for multi-access MIMO DDCSK-Walsh Coding-CD.

### IV. SIMULATION RESULTS

In this section, illustrative testing scenarios for the derived analytical expressions are plotted over a range of average SNR using different communications setups, Nakagami-*m* fading and various noise conditions. The results are also compared with the numerically evaluated expressions of the considered case studies. In the next plots, the solid lines are the results obtained from numerical evaluations, whereas the markers are generated from the derived analytical expressions. Moreover, we assume that $d_{SD} = d_{RD} = d_{SR} = 1$, $M = 32$ and $M_R=1$, unless stated otherwise, in the following four test scenarios.

In the first test, assuming AWGN environment (i.e. $a$=2) and $L$=$n$=2, and $M_D$=3, two cases are shown for the ABER for each of the EF and DF. In case 1, a Rayleigh fading (i.e. $m = 1$) is assumed for all links, while in case 2, Nakagami-*m* is assumed with $m = 4$. The results are shown in Fig. 5, where one can see the clear match between numerical and the derived EF and DF analytical ABER expressions. In the second test scenario, we assume the exact configurations as of the first scenario but with the assumption of Laplacian noise instead, and the results are shown in Fig. 6, where an excellent agreement can be seen between the approximate analytical and numerical results. The last two scenarios repeat the first two, while replacing the values of *L*, *n* and $M_D$ with 3, 3 and 4, respectively. The results are given in Fig. 7 and Fig. 8 and prove the accuracy and the validity of our derived results.

## V. CONCLUSION

In this paper, the performance analysis of the error rates in multi-access MIMO relay DCSK-CD system is considered for analysis. The analytical results are based on an approximation of the generalized $Q$-function newly developed to study the error rates in DCSK systems, which in a straightforward manner, also applies for DDCSK-Walsh Coding. The analysis assumed Nakagami-$m$ fading (which include Rayleigh fading as a particular case) and AWGGN environment (includes the Gaussian, the Laplacian, and other noise models as special cases). Numerical results showed that the derived analytical approximated expressions are very accurate. The proposed evaluation approach can also be extended in a very similar way to approximate the error rates and other performance metrics to many other chaos-based wireless communications systems. As a future work, we will study the usability of chaos communications in high mobility wireless networks such as VANETs [25] and the effect of interference and interference mitigation on such chaotic communications system [26].

## REFERENCES

[1] Y. Fang, G. Han, P. Chen, F. C. M. Lau, G. Chen and L. Wang, "A Survey on DCSK-based Communication Systems and Their Application to UWB Scenarios," *IEEE Communications Surveys & Tutorials,* pp. 1-34, March, 2016.

[2] E. Salahat, D. Shehada and C. Y. Yeun, "Novel Performance Analysis of Multi-Access MIMO Relay Cooperative RM-DCSK over Nakagami-m Fading Subject to AWGGN," in *IEEE 82nd Vehicular Technology Conference (VTC Fall)*, Boston, MA, 6-9 Sept. 2015.

[3] G. Kaddoum, "Design and Performance Analysis of a Multiuser OFDM Based Differential Chaos Shift Keying Communication System," *IEEE Trans. on Communications,* vol. 64, no. 1, pp. 249 - 260, Nov., 2015.

[4] H. Yang, W. K. S. Tang, G. Chen and G.-P. Jiang, "System Design and Performance Analysis of Orthogonal Multi-Level Differential Chaos Shift Keying Modulation Scheme," *IEEE Trans. on Circuits and Systems I: Regular Papers,* vol. 63, no. 1, pp. 146-156, Jan., 2016.

[5] F. J. Escribano, G. Kaddoum, A. Wagemakers and P. Giard, "Design of a New Differential Chaos-Shift Keying System for Continuous Mobility," *IEEE Trans. on Communications,* vol. 64, no. 5, pp. 2066-2078, Mar., 2016.

[6] U. Parlitz, L. O. Chua, L. Kocarev, K. S. Halle and A. Shang, "Transmission of Digital Signals by Chaotic Synchronization," *Int. J. Bifurcation and Chaos,* vol. 2, no. 4, p. 973–977, April, 1992.

[7] G. Kaddoum, E. Soujeri and Y. Nijsure, "Design of a Short Reference Noncoherent Chaos-Based Communication Systems," *IEEE Trans. on Communications,* vol. 64, no. 2, pp. 680 - 689, Jan., 2016.

[8] L. M. Pecora, T. L. Carroll and G. A. Johson, "Fundamentals of Synchronization in Chaotic Systems, Concepts, and Applications," *Int. J. Bifurcation Chaos,* vol. 74, p. 520–543, 1997.

[9] G. Kaddoum and E. Soujeri, "NR-DCSK: A Noise Reduction Differential Chaos Shift Keying System," *IEEE Trans. on Circuits and Systems II: Express Briefs,* vol. 63, no. 7, pp. 648 - 652, Feb., 2016.

[10] Y. Xia, C. K. Tse and F. C. M. Lau, "Performance of Differential Chaos-Shift-Keying Digital Communication Systems over a Multipath Fading Channel with Delay Spread," *IEEE Trans. on Circuits and Systems II: Express Briefs,* vol. 51, no. 12, p. 680–684, Dec, 2004.

[11] G. Kis, "Performance Analysis of Chaotic Communication Systems," Ph.D. Dissertation, Budapest University of Technology and Economics, Budapest, Hungary, Sept., 2005.

[12] H. Yang and G. P. Jiang, " High-Efficiency Differential-Chaos-Shift-Keying Scheme for Chaos-based Non-Coherent Communication," *IEEE Trans. on Circuits and Systems II: Express Briefs,* vol. 59, no. 5, p. 312–316, May, 2012.

[13] Y. Fang, J. Xu, L. Wang and G. Chen, "Performance of MIMO Relay DCSK-CD Systems over Nakagami Fading Channels," *IEEE Trans. on Circuits and Systems I: Regular Papers,* vol. 60, no. 3, p. 757–767, Mar., 2013.

[14] C. K. T. Yongxiang Xia and F. C. M. Lau, "Performance of Differential Chaos-Shift-Keying Digital Communication Systems Over a Multipath Fading Channel With Delay Spread," *IEEE Trans. on Circuits and Systems—II: Express Briefs,* vol. 51, no. 12, pp. 680-684, Dec., 2004.

[15] E. Salahat and A. Hakam, "Performance Analysis of Wireless Communications over α-η-μ and α-κ-μ Generalized Fading Channels," in *European Wireless Conference*, Barcelona, Spain, 14-16 May, 2014.

[16] P. Chen, L. Wang and G. Chen, "DDCSK-Walsh Coding: A Reliable Chaotic Modulation-Based Transmission Technique," *IEEE Trans. on Circuits and Systems—II: Express Briefs,* vol. 59, no. 2, pp. 128-132, Feb., 2012.

[17] E. Salahat and A. Hakam, "Novel Unified Expressions for Error Rates and Ergodic Channel Capacity Analysis over Generalized Fading Subject to AWGGN," in *IEEE Global Communication Conference*, Austin, U.S.A, 8-12 Dec., 2014.

[18] E. Salahat and H. Saleh, "Novel Average Bit Error Rate Analysis of Generalized Fading Channels Subject to Additive White Generalized Gaussian Noise," in *IEEE Global Conference on Signal and Information Processing*, Atlanta, USA, 3-5 Dec. 2014.

[19] E. Salahat and I. Abualhoul, "Generalized Average BER Expression for SC and MRC Reciever over Nakagami-m Fading Channels," in *IEEE International Symposium on Personal, Indoor and Mobile Radio Communications*, London, UK, 8-11 Sept., 2013.

[20] E. Salahat and I. Abualhaol, "General BER Analysis over Nakagami-m Fading Channels," in *Joint IFIP Wireless and Mobile Networking Conference (WMNC)*, Dubai, 23-25 April 2013.

[21] P. Moschopoulos, "The Distribution of the Sum of Independent Gamma Random Variables," *Annals Inst. Statistical Mathematics,* vol. 37, pp. 541-544, 1985.

[22] E. Salahat, "Unified Performance Analysis of Maximal Ratio Combining in η-μ, λ-μ and κ-μ Generalized Fading Channels," in *IEEE 80th Vehicular Technology Conference (VTC'14)*, Vancouver, Canada, 14-17 Sept. 2014.

[23] D. Z. Alan Jeffrey, Table of Integrals, Series, and Products, 7th Edition, Academic, 2007.

[24] M. Ding, Multi-point Cooperative Communication Systems: Theory and Applications, Springer, 2013.

[25] L. Bariah, D. Shehada, E. Salahat and C. Y. Yeun, "Recent Advances in VANET Security: A Survey," in *IEEE 82nd Vehicular Technology Conference (VTC Fall)*, Boston, MA, 6-9 Sept. 2015.

[26] A. Hakam, R. Shubair, S. Jimaa and E. Salahat, "Robust Interference Suppression using a new LMS-based Adaptive Beamforming Algorithm," in *IEEE Mediterranean Electrotechnical Conference*, Beirut, 13-16 April 2014.